\documentclass[twocolumn,showpacs,preprintnumbers,amsmath,amssymb,floatfix]{revtex4}
\usepackage{graphicx}
\usepackage{dcolumn}
\usepackage{bm}
\begin{document}


\title[Dark matter accretion ]{Dark Matter Accretion into
  Supermassive Black Holes}

\author{S\'ebastien Peirani}
\affiliation{Institut d'Astrophysique de Paris, 98 bis Bd Arago,
 75014 Paris, France - Unit\'e mixte de 
         recherche 7095 CNRS - Universit\'e
         Pierre et Marie Curie}
\email{peirani@iap.fr}
\author{J.A. de Freitas Pacheco}
\affiliation{Observatoire de la C\^ote d'Azur\\
Laboratoire Cassiop\'ee - BP 4229 - 06304 - Nice Cedex 4 - France}
\email{pacheco@obs-nice.fr}
\date{\today}






\begin{abstract}

The relativistic accretion rate of dark matter by a black hole is revisited. Under
the assumption that the phase space density indicator, $Q=\rho_{\infty}/\sigma^3_{\infty}$,
remains constant during the inflow, the derived accretion rate can be higher up to
five orders of magnitude than the classical accretion formula, valid for non-relativistic
and non-interacting particles, when typical dark halo conditions are considered. For
these typical conditions, the critical point of the flow is located at distances of 
about 30-150 times the horizon radius. Application of our results to black hole seeds
hosted by halos issued from cosmological simulations indicate that dark matter contributes
to no more than $\sim$ 10\% of the total accreted mass, confirming that the bolometric
quasar luminosity is related to the baryonic accretion history of the black hole.  

\end{abstract}



\maketitle

\section{Introduction}

The presence of supermassive black holes (SMBHs) in the center either of elliptical galaxies
or bulges of spiral galaxies seems presently to be a well established fact \cite{kor05}.
However, in spite of a large number of studies performed in the past years, the origin and 
evolution of these objects remain quite uncertain. Since SMBHs cannot be formed as a consequence
of the evolution of stars having comparable mass, they are generally supposed to evolve from
``seeds'', which grow by accretion inside the host galaxy. Moreover, the observation of quasars already
at $z \sim$~6.5, when the universe was only $\sim$~0.85~Gyr old, implies that the process of
growth (accretion) must have been quite efficient at those early times. 

Seeds could be intermediate mass ($10^3$-$10^4$~$M_{\odot}$) black holes formed in the collapse 
of primordial gas clouds \cite{hae93,eis95,kou04} or in the core collapse of relativistic 
star clusters formed in starbursts, which may have occurred in the early evolution 
of galaxies \cite{sha04}. Another appealing possibility is the formation of black hole seeds 
with masses in the range 100-300~$M_{\odot}$, resulting from the evolution of primordial massive stars
\cite{heg03}. These stars would have been formed at $z \sim$~15-20 in the high density peaks of
the primordial fluctuation spectrum and their high masses would be a consequence of
a very inefficient gas cooling at zero metallicity \cite{car84}. Besides forming black hole seeds,
these massive stars could also have been the source of UV photons, responsible for the reionization of
the universe.

The analysis of the bolometric quasar luminosity function at different redshifts indicates that
most of the accreted mass by black holes should be baryonic in origin \cite{sol82,sma92,hop06}. Although 
expected to be small, the contribution of dark matter to the black hole growth is 
still uncertain. Unlike baryons, the infall of dark matter into black holes does not 
produce electromagnetic radiation and thus, does not contribute to the quasar luminosity, although
some energetic photons could be produced either by annihilation or decay, if dark matter is constituted
by neutralinos or sterile neutrinos respectively. Moreover, dark 
matter particles constitute a collisionless fluid and, consequently, the accretion process should 
be less efficient than that expected for a dissipative fluid. Even so, different scenarios were
envisaged in which dark matter could give a significant contribution to the mass accretion budget. If halos
embedding galaxies are constituted by self-interacting dark matter \cite{dav01}, then black holes with
masses $\geq 10^6~M_{\odot}$ could be formed directly as a consequence of relativistic core collapse of
halos, following the gravothermal catastrophe \cite{bal02}. In this scenario, no baryons and no prior
star formation are required to form the massive black holes, but the high interaction cross-section
necessary for the mechanism to work remains a difficulty. A different approach was considered in
\cite{zel03}, where the dark matter accretion process was investigated in a scenario in which stars
are already formed in the core of the galaxy. Dark matter particles scatter on stars in the vicinity
of the central black hole, producing an important inflow as a consequence of diffusion in phase space and
a mass growth scaling as $M_{bh} \propto t^{9/16}$. A similar mechanism was examined in \cite{mun04} but,
in this case, dark matter particles are scattered by molecular clouds. According to the authors, this
process can boost the growth of the mass of seeds from about $\sim 5~M_{\odot}$ up 
to $\sim 10^4~M_{\odot}$, which
then grow by Eddington limited baryonic accretion. The growth of seeds under adiabatic conditions was 
considered by \cite{mac02}, who obtained a mass-age
relation only in the isothermal case but with the resulting mass-velocity dispersion relation
differing from observations. However, they have obtained a better agreement with 
data by assuming that black holes grow proportionally to the host halo mass.

In the present work, a new investigation on the dark matter accretion process into black holes is
reported. Our approach gives an accretion rate significantly higher than that based on the well-known
cross-section for capture by a black hole of non-relativistic and non-interacting particles. We have performed 
some numerical applications of our formula to the growth of seeds inside dark matter halos issued from 
cosmological simulations. These numerical calculations indicate that, if there is a coeval growth of
seeds and their host halos, then dark matter contributes to no more than $\sim$~10\% of the total accreted 
matter, giving a further support to the usual assumption that the bolometric quasar luminosity function is
related to the baryonic accretion history of SMBHs. This paper is organized
as follows: in Section II, the equations describing the accretion process are presented; in
Section III, numerical examples are discussed and finally, in Section IV, our main conclusions are
given.

\section{The accretion model}

The relativistic steady spherical flow of a perfect gas into a black hole was first investigated by
\cite{mic72} and later by \cite{beg78,bri80,das01,man07} among others. Radiation transfer effects were
considered in the pioneering work by Thorne and collaborators (\cite{tho81,fla82}) and more recently in
\cite{nob91,par06}. A comprehensive review on the subject can be found in \cite{cha96}. 
It is worth mentioning that if the fluid has a non-zero angular momentum with respect to the black
hole, then the spherical symmetry is broken. In this case, the accretion process is more complex since mechanisms
of angular momentum transfer must be considered. Studies by \cite{macmillan04} based on a 
``shell" accretion model indicate that the black hole growth factor is reduced if the fluid has
a non-zero angular momentum. As we shall see later, our cosmological simulations do not have
enough resolution to describe adequately the central region including the black hole
influence radius. Thus, in the present investigation, we restrict our analysis to the spherical
case. 

Past investigations have shown that for an adiabatic flow of a gas with $\gamma < 5/3$, the only 
critical point of the flow which lies outside the horizon is the one corresponding to the Bondi solution. As we
shall see below, a similar situation occurs in the case of a collisionless fluid.
In our scenario, seeds appear around $z \sim 12-15$ as a result of the core-collapse
of massive stars \cite{heg03} and are located in the center of dark matter halos.
Near the central black hole of mass $M_{bh}$, we assume that the spacetime is described 
by the Schwarzschild metric and that a steady spherical inflow is set up inside the influence 
radius of the black hole, defined by $r_* = 2GM_{bh}/\sigma^2_{\infty}$, where 
$\sigma_{\infty}$ is the velocity dispersion of
dark matter particles of mass $m$ in the core of the halo.

Under these conditions, the equations of motion are: the conservation of
the mass flux
\begin{equation}
J^k_{;k} = 0
\label{massconservation}
\end{equation}
and the energy-momentum flux conservation
\begin{equation}
T^k_{i;k} = 0
\end{equation}
where $J^k = mnu^k$ is the mass-current density and the energy-momentum tensor is that of an ideal fluid, i.e.,
$T^k_i = (P+\varepsilon)u^ku_i-P\delta^k_i$, with $P$, $\varepsilon$, $n$, being respectively 
the proper pressure, the proper energy density and the proper number density 
of the dark matter fluid. $u^i = dx^i/ds$ is the 4-vector velocity, whose non-null 
components are $u \equiv u^1 = dr/ds$ and $u^0 = dt/ds$. As in \cite{mic72}, one obtains 
from these equations
\begin{equation}
\frac{(P+\varepsilon)}{n}(1-\frac{r_g}{r}+u^2)^{1/2} = k_1
\label{pressure}
\end{equation}
where $r_g = 2GM_{bh}/c^2$ is the gravitational radius and $k_1$ is a constant.
Deriving eq.~\ref{pressure} with respect to the radial coordinate $r$ and using the mass conservation 
equation (eq.~\ref{massconservation}), one obtains a ``wind'' equation, i.e.,
\begin{equation}
\frac{dlg~u}{dlg~r}\lbrack u^2-V^2F(r,u)\rbrack+\lbrack\frac{r_g}{2r}-2V^2F(r,u)\rbrack=0
\label{wind}
\end{equation}
where
\begin{equation} 
F(r,u)=(1-\frac{r_g}{r}+u^2)
\end{equation}
and
\begin{equation}
V^2 = \frac{dlg~(P+\varepsilon)}{dlg~n}-1
\label{sound}
\end{equation}

The critical point of the flow occurs where both bracketed factors in eq.~\ref{wind} vanish
simultaneously, namely,
\begin{equation}
u_c^2 = \frac{r_g}{4r_c}
\label{criticalradius}
\end{equation}
and
\begin{equation}
V^2_c = \frac{u^2_c}{(1-3u^2_c)}
\label{soundcritical}
\end{equation}
 
After decoupling, dark matter particles constitute a weakly interacting fluid. In the matter era,
at higher redshifts, the universe evolves closely to the Einstein-de Sitter 
model. In this case, the distribution function
$f(\vec u,t)$ of dark matter particles is a function of integrals of the characteristics of the
Vlasov equation, i.e.,
\begin{equation}
f(\vec u,t) = f(a(t)\vec u)
\end{equation}
where $a(t) \propto t^{2/3}$ is the scale factor and $\vec u =(\vec v - H\vec r)$ is the
peculiar velocity. The distribution function must also satisfy the normalization condition
\begin{equation}
\int f(\vec u,t)d^3\vec u = mn(t) = \rho(t) = \frac{1}{6\pi Gt^2} \propto a^{-3}(t)
\end{equation}
where the small contribution of baryons to the total matter density was neglected. Under 
these conditions, the velocity dispersion of the peculiar velocity satisfies
\begin{equation}
\sigma^2 = \frac{\int u^2f(\vec u,t)d^3\vec u}{\int f(\vec u,t)d^3\vec u}\propto \frac{1}{a^2(t)}
\label{dispersion}
\end{equation}
and the quantity $Q = \rho(t)/<\sigma^2>^{3/2}$, an indicator of the phase space density,
is conserved during the expansion of the universe. 

High resolution simulations of galaxy-size dark matter
halos by \cite{taylor01} indicate an increase of Q toward the center. Similar results were obtained
by \cite{rasia03}, who have derived a power-law variation, e.g., $Q\propto r^{-\beta}$ for 
cluster-size halos, with $\beta$ quite close to the value found by \cite{taylor01}, namely
$\beta \simeq 1.87$. No adequate explanation exists presently for such a power-law profile, although
\cite{henriksen06} suggests that it could be a consequence of an isotropic distribution function
produced in the halo core by processes of relaxation related to the radial orbit instability. In fact,
\cite{henriksen06,henriksen07} assuming that the evolution of halos is self-similar, derived the
conditions required for the density profile as well as the phase density indicator $Q$ be described by 
power laws. A density profile with a slope $dlg\rho/dlgr=-2$, typical of self-similar radial infall, can
be consistent with phase space profiles either of the form $Q=\rho/\sigma^3$ or $Q=\rho/\sigma^2$.
Here we adopt the usual definition, which is supported by the results of different cosmological simulations.
 
As structures begin to form as well
as to accrete matter and merger together, rapid variations in the gravitational potential  contribute to
the relaxation of the system through ``violent relaxation'' (\cite{lyn67}). This process leads
to a smooth mass-independent distribution function as a result of the gravitational
scattering of particles. Violent relaxation produces a more mixed system in phase space,
reducing the value of the coarse-grained distribution function or, in other words,
reducing the value of $Q$ (\cite{tre86}). Therefore, as a consequence of such a mechanism, the
velocity dispersion of dark matter particles, at late epochs, inside halos is not negligible as one 
would expect taking into account only the expansion of the universe (eq.~\ref{dispersion}). In this 
case, including the energy of the
random peculiar motion, the total energy density can be written as
\begin{equation}
\varepsilon = mnc^2 + \frac{1}{2}mn\sigma^2
\end{equation}
and, consequently
\begin{equation}
(P + \varepsilon) = mnc^2 + \frac{5}{6}mn\sigma^2
\label{pressure2}
\end{equation}

 Notice that, as mentioned previously, the phase space density increases toward the center as
$Q\propto r^{-1.87}$, but one expects a convergence to a finite value close to the black hole
influence radius. Once the inflow develops, we hypothesize that the
quantity $Q$ remains constant. This assumption is reasonable if no important phase
space mixing occurs during the inflow.
According to cosmological simulations by \cite{pei06,pei07}, the phase space indicator 
$Q$ estimated in the core of halos evolves in two distinct phases. An early 
and fast phase in which the decrease of $Q$ is
associated to virialization just after the first shell crossing, following the gravitational
collapse of the halo. In a second late and long phase, a slow decrease of $Q$ is observed, consequence
of continuous matter accretion and merger episodes. Since the infall time scale of dark matter particles
onto the central black hole is small
in comparison with the characteristic time scale in which the phase space indicator varies,
namely, $Q/\mid (dQ/dt)\mid \sim$ 0.7 Gyr (typical value derived from the aforementioned simulations), we 
can suppose that the flow is steady but adjusted to the slowly varying conditions 
of the environment. In this case, eq.~\ref{pressure2} can be recast as
\begin{equation}
(P + \varepsilon) = mnc^2 + \frac{5}{6}m\frac{n^{5/3}}{Q_n^{2/3}}
\label{pressure3}
\end{equation}
Notice that, in the above equation, we have defined $Q_n = Q/m = n/\sigma^3$ and that the
assumption of keeping $Q$ constant during the inflow is similar to that of considering
the compression of a perfect gas with $\gamma = 5/3$. In fact, since the entropy density 
satisfies $s \propto -log~Q$, the assumption that $Q$ remains constant is equivalent to the hypothesis 
of an adiabatic inflow, although with a different adiabatic constant.

Using eq.~\ref{pressure3}, one obtains
\begin{eqnarray}
\frac{d(P+\varepsilon)}{dn}= mc^2 + \frac{25}{18}m(\frac{n}{Q_n})^{2/3}\nonumber\\
 = \frac{(P+\varepsilon)}{n}\lbrack 1 + \frac{5}{9}\frac{mn^{5/3}Q_n^{-2/3}}{(P+\varepsilon)}\rbrack
\end{eqnarray}
and replacing the equation above into eq.~\ref{sound}, one obtains
\begin{equation}
V^2 \simeq \frac{5}{9}\frac{n^{2/3}Q_n^{-2/3}}{c^2}-\frac{25}{54}\frac{n^{4/3}Q_n^{-4/3}}{c^4}
\label{soundvelocity}
\end{equation}
where a Taylor series up to the second order was used. The constant $k_1$ in eq.~\ref{pressure} can be
calculated by imposing that for $r >> r_g$ we have $u \rightarrow 0$. Thus, one obtains
\begin{equation}
k_1 = mc^2 + \frac{5}{6}m(\frac{n_{\infty}}{Q_n})^{2/3} 
\end{equation}
where $n_{\infty}$ stands for the dark matter particle number density far away from the black 
hole, where the flow velocity is negligible. At the critical point, eq.~\ref{pressure} can be 
rewritten as
\begin{eqnarray}
\lbrack mc^2+\frac{5}{6}m(\frac{n_c}{Q_n})^{2/3}\rbrack(1-3V_c^2)^{1/2} =\nonumber\\
= mc^2+\frac{5}{6}m(\frac{n_{\infty}}{Q_n})^{2/3}
\end{eqnarray}
where $n_c$ is the number density at the critical point. Replacing eq.~\ref{soundvelocity} into the
equation above, one obtains after some algebra,
\begin{equation}
\frac{n_c}{n_{\infty}} \simeq \frac{6}{5}(\frac{c}{\sigma_{\infty}})^2
\end{equation}
which permits to estimate the density at the critical point as a function of physical conditions of the halo,
far away the black hole. Since the flow velocity at the critical 
point satisfies $u_c^2<<1/3$,
from eq.~\ref{soundcritical} and eq.~\ref{soundvelocity} one obtains to the first order 
\begin{equation}
u_c^2 \simeq \frac{5}{9}\frac{n_c^{2/3}}{Q^{2/3}c^2}
\end{equation}
and, using the relation above for the density at the critical point and the fact 
that $Q_n=n_{\infty}/\sigma_{\infty}^3$, one obtains finally
\begin{equation}
u_c^2 \simeq 0.627(\frac{\sigma_{\infty}}{c})^{2/3}
\end{equation}
Replacing this result into eq.~\ref{criticalradius}, one obtains for the critical radius
\begin{equation}
r_c \simeq 0.398(\frac{c}{\sigma_{\infty}})^{2/3}r_g \simeq 0.398(\frac{\sigma_{\infty}}{c})^{4/3}r_* 
\end{equation}
For typical halos with $\sigma_{\infty}$ in the range 50-300 km/s, the critical radius is located in 
the interval 130-40 times the horizon radius, well inside the influence sphere. It is interesting
to recall that, for an ideal gas, the critical point is located at 
\begin{equation}
r_c \simeq \frac{(5-3\gamma)}{8}(\frac{c}{a_{\infty}})^2
\end{equation}
where $a_{\infty}$ is the sound velocity far away from the black hole. This relation implies that
the critical point is outside the horizon only if $\gamma < 5/3$ (see \cite{beg78}).

Once the parameters at the critical point are known, the accretion rate is given by
\begin{eqnarray}
\frac{dM_{bh}}{dt} = 4\pi r_c^2(\frac{T^1_0}{c})\nonumber\\
 = \frac{27\pi}{4\sqrt{125}}r_g^2c^4Q = \frac{27\pi}{\sqrt{125}}(GM_{bh})^2Q
\label{accretion}
\end{eqnarray}
where only first order terms were retained. Notice that for non-relativistic and
non-interacting particles, the expected accretion rate is \cite{zel71}
\begin{equation}
\frac{dM_{bh}}{dt}=4\pi r_g^2\rho_{\infty}\frac{c^2}{v_{\infty}}
\end{equation}
where $v_{\infty}$ is the relative velocity of particles ``at infinity"  with respect 
to the black hole. This equation predicts an
accretion rate smaller by a factor $27c^2/(16\sqrt{125}\sigma_{\infty}^2)$ than eq.~\ref{accretion},
which is of the order $10^5$, if estimated in the core of a typical dark halo.

\section{Numerical application}

In this section, an application of the results obtained previously is performed. Firstly, we
emphasize that we do not intend to present here a detailed model for the growth of black holes
located in the core of halos. This problem will be discussed in a future paper. Our purpose here is to give 
only a rough estimate of the dark matter fraction accreted by the black hole, assuming that
our formula (eq.~\ref{accretion}) gives a correct description of the inflow.

Halos investigated in this work were selected from catalogs prepared by \cite{pei06}, which were 
based on systems identified in cosmological N-body simulations, using an adaptive particle-particle/particle-mesh
($AP^3M$) code (\cite{cou95}). The adopted cosmological parameters were $h = 0.65$, $\Omega_m = 0.3$
and $\Omega_{\Lambda} = 0.7$, with the power spectrum normalization taken to be $\sigma_8 = 0.9$.
Simulations were performed in a box of side $30h^{-1}~Mpc$, including $256^3$ particles, corresponding
to a mass resolution of $2.05\times 10^8~M_{\odot}$ and covering the redshift interval $0 \leq z \leq 50$.
Halos were initially detected by using a friends-of-friends (FOF) algorithm (\cite{dav85}) and, in a
second step, unbound particles were removed by an iterative procedure. In the total, about 9000 gravitationally
bound halos were detected at $z=0$. For more details, see \cite{pei06,pei04}. However, the mass history in the redshift
range $0 \leq z \leq 10$ could be followed only for 35 halos with enough particles ($N \geq 50$).
The phase space density indicator $Q$, was calculated according to the prescription
detailed in \cite{pei06}, inside a spherical volume whose radius is equal to one tenth of the gravitational
radius of the halo, which is defined as the ``core" region. 

The present observations indicate that in all massive galaxies where the masses of bulges and central
black holes have been estimated from kinematic data, there is a robust correlation between them
\cite{geb00,tre02,onk04}. Moreover, other investigations have shown conclusively that not 
only there is a link between the black hole mass and the bulge mass but also with the circular
velocity \cite{baes03}. If circular motions probe the halo potential, then a relation between the
present black hole mass and the present halo mass is also expected. These relations suggest that
black holes grew coevally with their host galaxies and associated halos \cite{mil06}. Here, this
 assumption will be adopted and, consequently, the black hole growth rate is given by
\begin{equation}
\frac{1}{Y(t)}\frac{dY(t)}{dt} = \alpha\frac{dlg~M_{H}(t)}{dt}
\label{growth1}
\end{equation}
where $\alpha$=1.27 is a constant (see, eq.~\ref{baes} below) and $Y(t)=M_{bh}(t)/M_{bh}(0)$ is the black hole 
mass, measured in units of its present mass $M_{bh}(0)$. The right member of eq.~\ref{growth1} represents 
the mass growth rate of the host dark halo.

At any instant, the total mass of the black hole is the sum of the amount of accreted baryonic and dark matter
masses. If $X(t)$ is fraction of accreted dark matter at instant $t$ then, using eq.~\ref{accretion} and the
previous definitions, one obtains
\begin{equation}
\frac{dX(t)}{dt}=-\frac{X(t)}{Y(t)}\frac{dY(t)}{dt}+KY(t)Q(t)
\label{growth2}
\end{equation}
where, as in eq.~\ref{growth1}, the time is in units of $H_0^{-1}$ and the constant $K$ is equal to
$K=27\pi G^2M_{bh}(0)/\sqrt{125}H_0=2.143(M_{bh}(0)/M_{\odot})$. The numerical value corresponds to
the phase space density taken in units of $M_{\odot}pc^{-3}km^{-3}s^{3}$.

The calculation procedure was the following: for a given halo having presently a mass $M_H$, the 
black hole mass was estimated by the relation derived by
\cite{baes03}, namely
\begin{equation}
M_{bh}(0) = 1.1\times 10^7(\frac{M_H}{10^{12}M_{\odot}})^{1.27}~M_{\odot}
\label{baes}
\end{equation}
and, consequently, the value of the constant K can be calculated. The system of 
differential eqs.~\ref{growth1} and \ref{growth2}
was then solved numerically, using the mass and phase space density derived from cosmological simulations.
The initial values of $Y$ and $X$ were varied until the solution gives $Y=1$ for the present black hole mass.

For studied halos, with present masses in the range $7.5\times 10^{11} - 9.54\times 10^{12}~M_{\odot}$,
we have found that the fraction of dark matter accreted by the halo is in the range 1-10\%.

\begin{figure}
\includegraphics[width=\columnwidth]{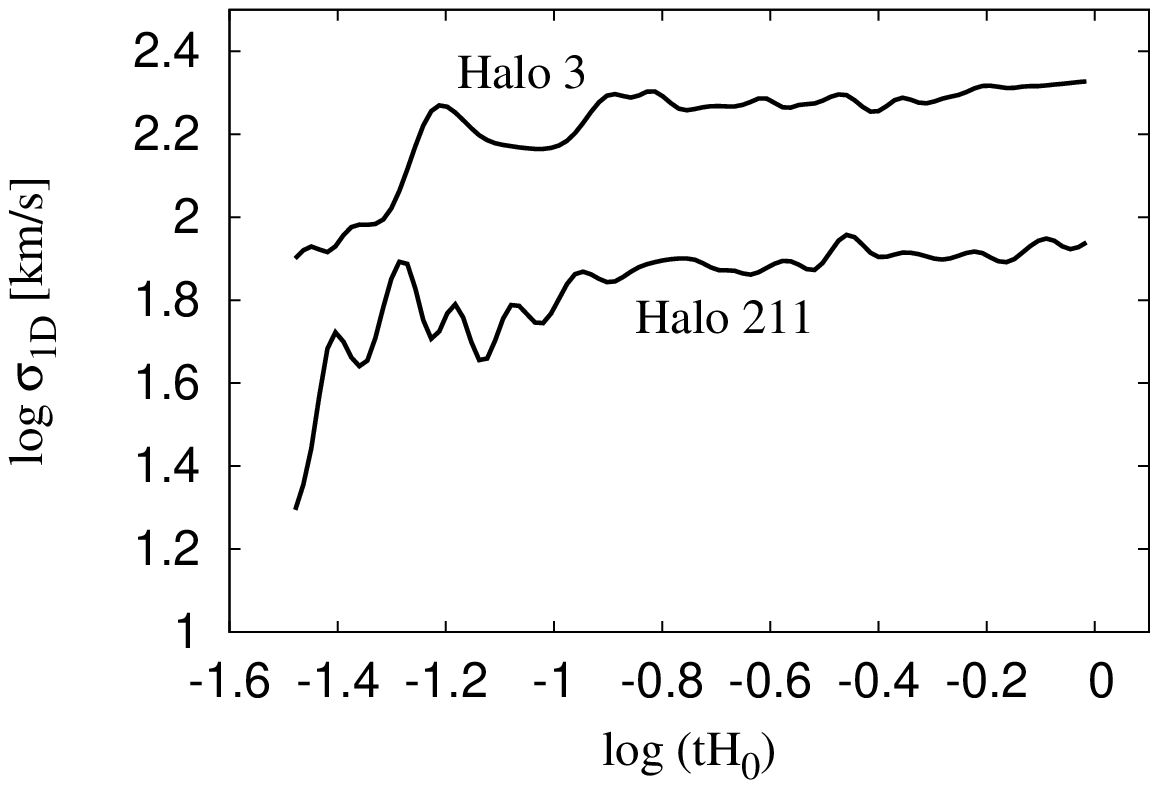}
\includegraphics[width=\columnwidth]{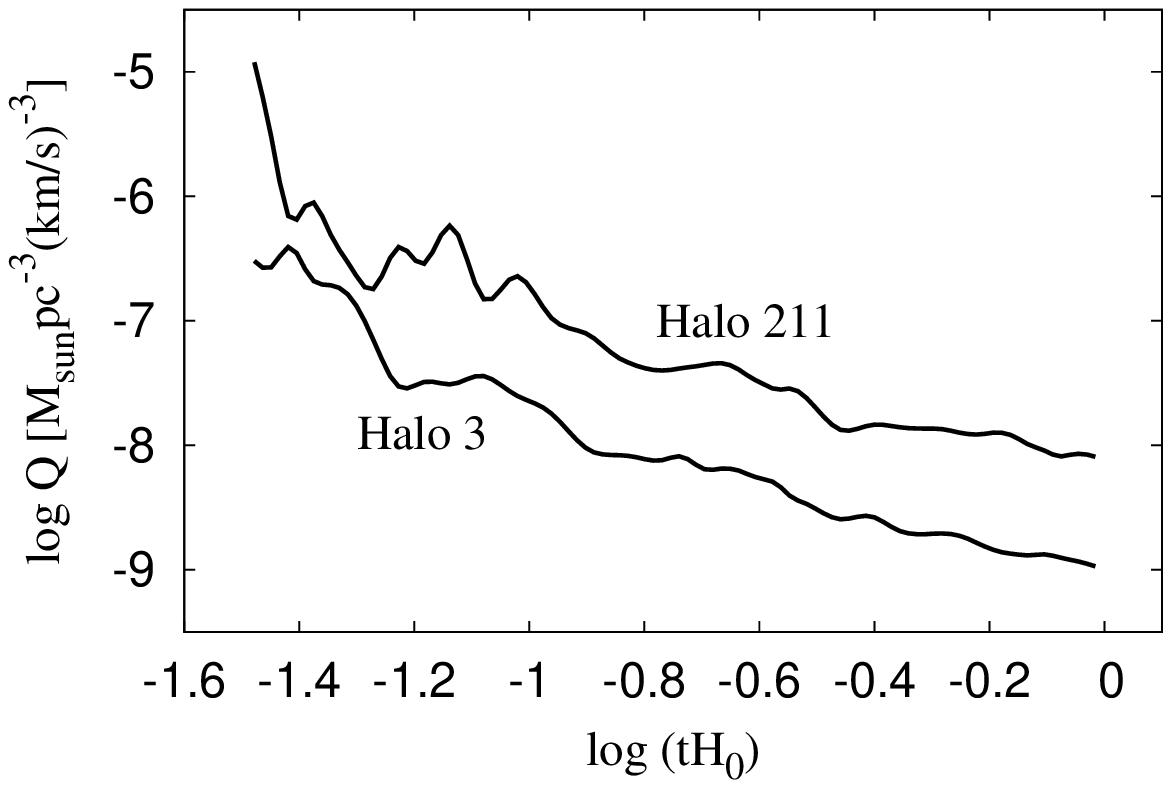}
\caption{Evolution of the velocity dispersion (km/s) inside the core of halos 3 and 211 (upper panel).
Evolution of the phase density ($M_{\odot}pc^{-3}km^{-3}s^3$) in core for the same halos (lower panel).
Time is given in units of $H_0^{-1}$.}
\end{figure}

In order to exemplify, in figure 1 it is shown the evolution of two halos whose final masses are respectively
$7.46\times 10^{11} M_{\odot}$ (halo 211) and $8.06\times 10^{12} M_{\odot}$ (halo 3). The upper panel shows
the evolution of the velocity dispersion whereas the lower panel shows the evolution of the phase space density
indicator $Q$. Both quantities were calculated in the core region, as defined above. Notice that to make 
easier the comparison with observations, we have plotted the 1D-velocity dispersion and that the quantity $Q$ was 
also calculated with respect to $\sigma_{1D}$. The 3D values can be obtained multiplying the 
1D-velocity dispersion by $\sqrt{3}$ and dividing $Q_{1D}$ by $\sqrt{27}$. Notice also that the more
massive halo (halo 3) has a higher velocity dispersion but a lower phase density $Q$, in agreement
with the fact that the late quantity, for a given redshift, scales as $Q \propto M^{-0.82}$ (\cite{pei06}). The 
black hole mass evolution in the considered halos is shown in figure 2 (upper panel), where it is possible
to see that the growth is not uniform, retracing the mass history of halos. Sudden increases in mass growth
represent merger episodes. The fraction of accreted dark matter is shown in the lower panel of figure 2.
At $z \sim 10$, the fraction of dark matter is about $0.005$ for both halos whereas it attains a value
around $0.02$ for the less massive halo (halo 211) and about $0.05$ for halo 3. Notice the fast increase
in the dark matter contribution for $z < 2$, consequence of a decreasing merger rate after such a redshift.
This affects the dark matter accretion in two aspects: firstly, decreasing the contribution of baryonic matter
to the black hole growth and secondly, the decreasing merger rate implies also that $Q$ decreases more slowly
than the black hole growth rate and, consequently, the relative contribution of dark matter increases. 

\begin{figure}
\includegraphics[width=\columnwidth]{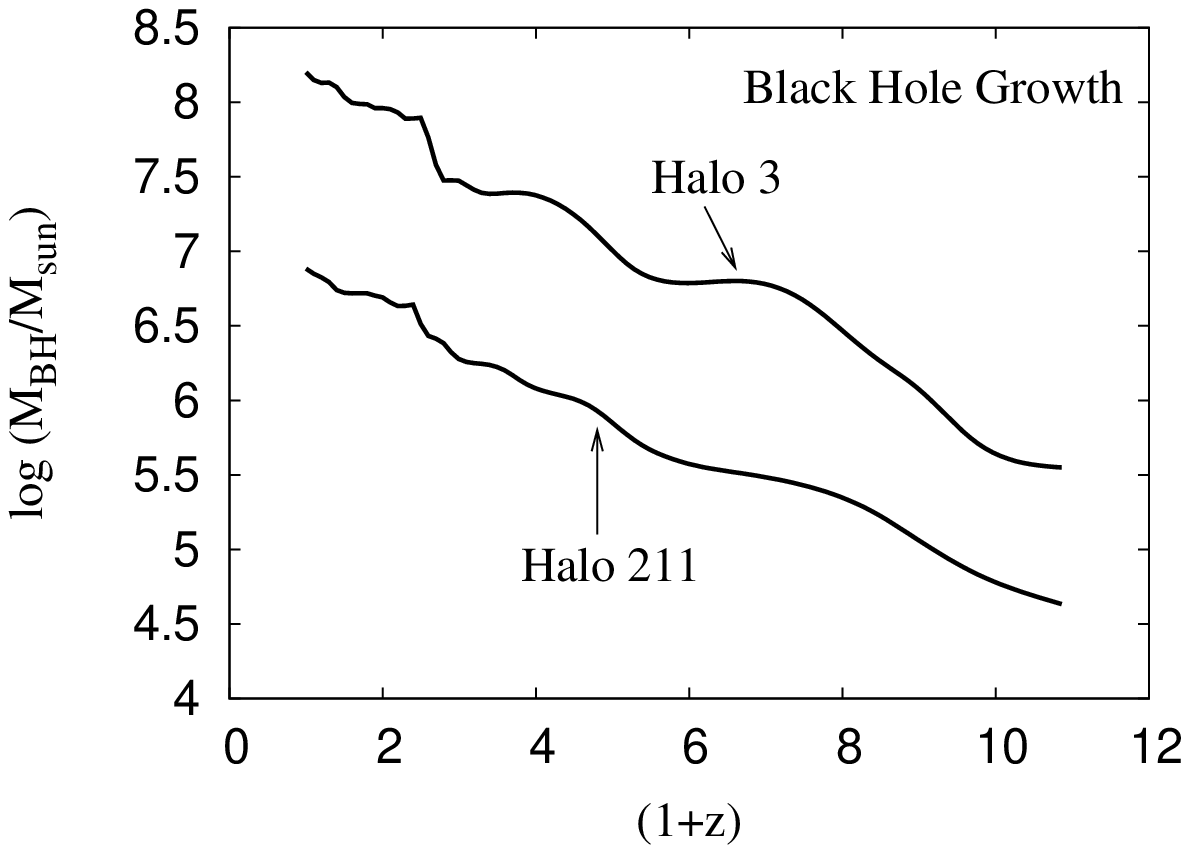}
\includegraphics[width=\columnwidth]{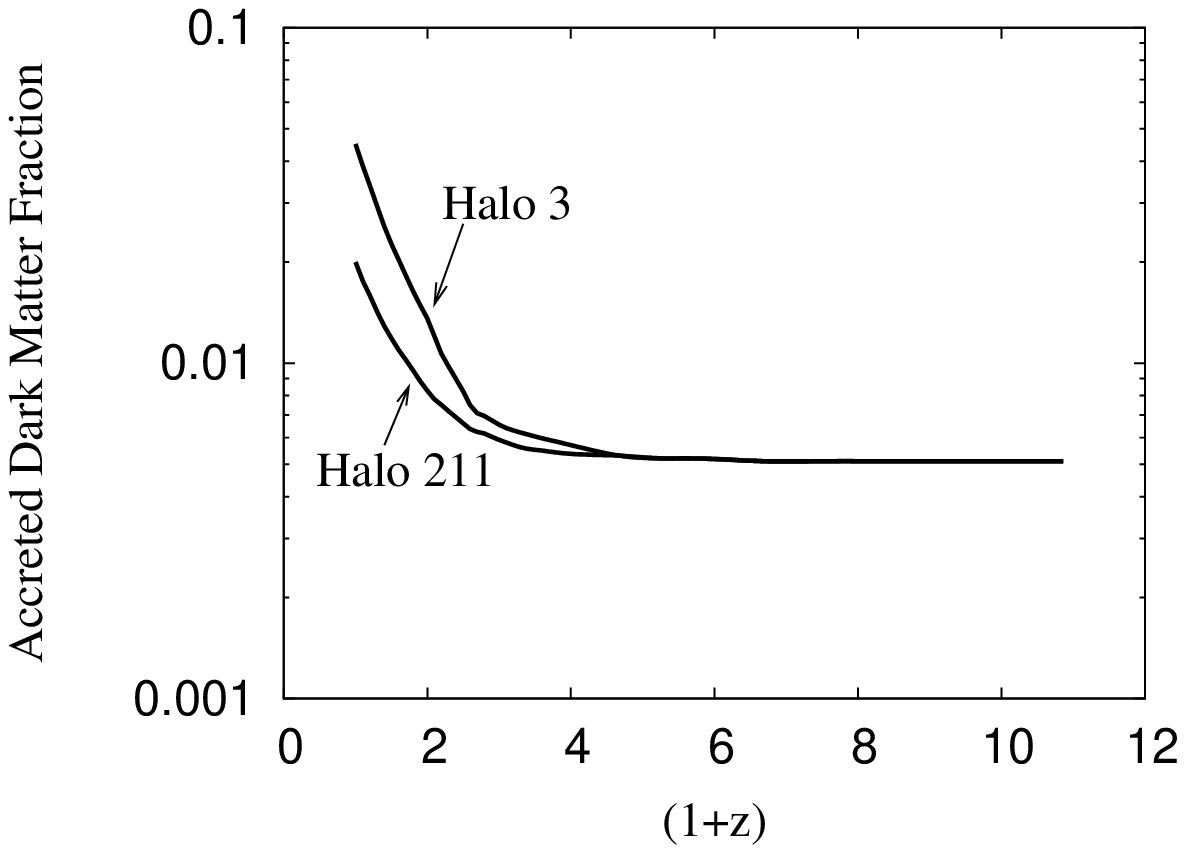}
\caption{In the upper panel it is shown the mass evolution of black holes while in the lower panel
the evolution of the fraction of accreted dark matter is shown.}
\end{figure}

In figure 3 the baryonic and dark matter accretion rates are compared for both halos. Different peaks 
in the baryonic accretion rate are associated to minor or major merger episodes, with the maximum rate
occurring around $z \sim 1.5$. Halo 3 is an example where for two short time intervals the
accretion rate of dark matter became higher than the baryonic accretion rate. The baryonic accretion
rate will be probably reduced by a factor of 2-6 if the star formation process is included, an
aspect presently under investigation, based on detailed cosmological simulations.

\begin{figure}
\includegraphics[width=\columnwidth]{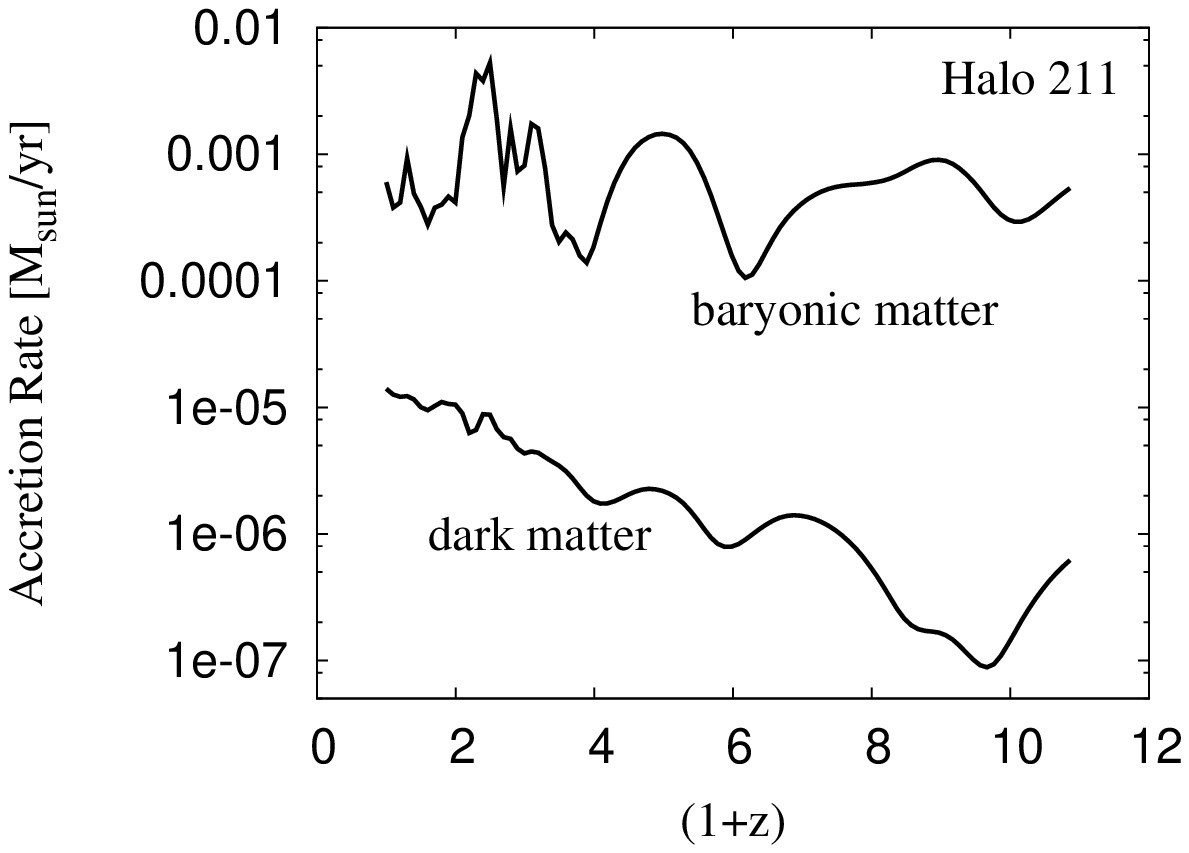}
\includegraphics[width=\columnwidth]{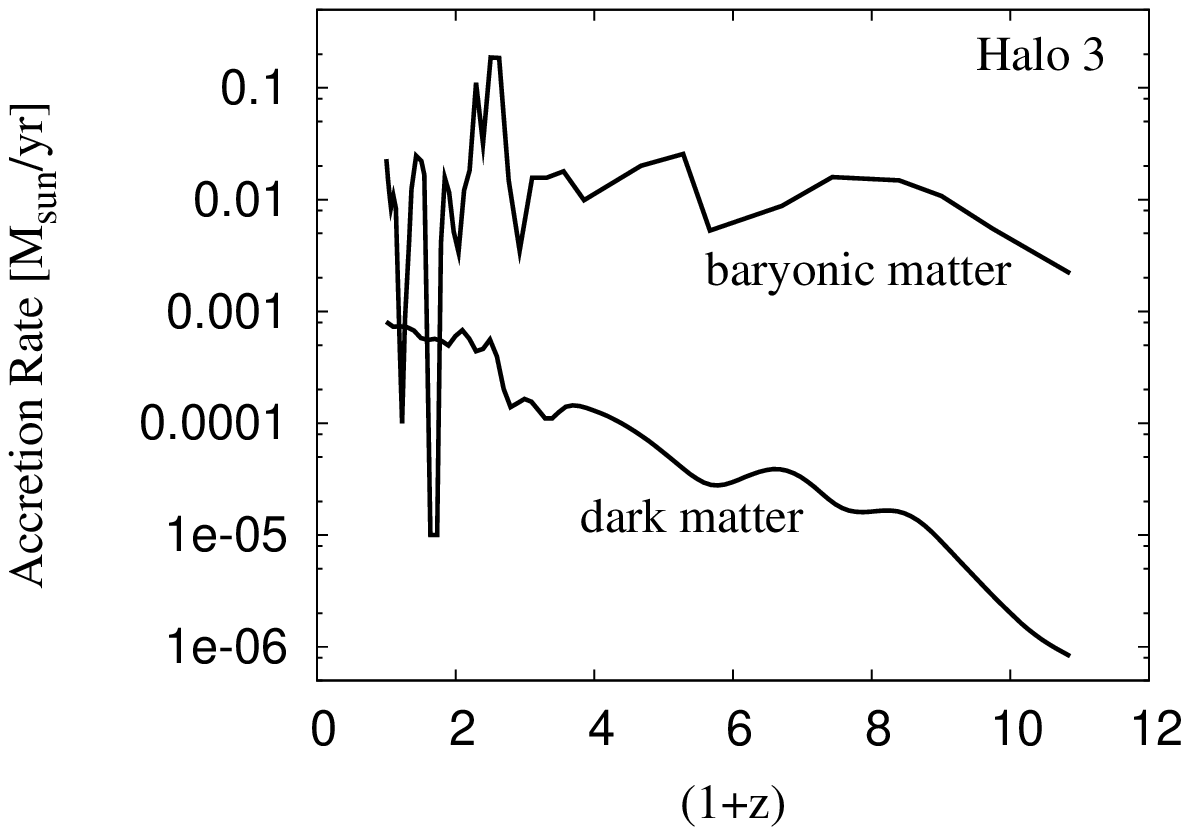}
\caption{Evolution of the baryonic and dark matter accretion rates for halo 211 (upper panel) and
halo 3 (lower panel).}
\end{figure}

\section{Conclusions}

The accretion of dark matter into black holes located in the center of galaxies and their halos
was studied. The accretion rate was derived by solving the equations for a relativistic steady
spherical symmetric inflow, under the assumption that the phase space density indicator 
$Q = \rho/\sigma^3$ remains constant. This hypothesis is equivalent to say that the
inflow is adiabatic, since the entropy density satisfies $s \propto - log~Q$. 

We found that our solution predicts that the critical radius is always outside the black hole
horizon and that the flow radial velocity at the critical point is of the order of few
percent of the speed of light. After the crossing the critical point, numerical solutions
of equations describing the flow indicate that the velocity and the density have power-law
profiles, e.g., $u \propto 1/r^{0.6}$ and $\rho \propto 1/r^{1.4}$, a behavior quite close to 
that resulting from the classical ``free-fall" solution. Notice that the growth of black holes
embedded in a collisionless fluid was also studied by considering that orbits shrink inside
the influence sphere but conserving their adiabatic invariants \cite{you80}. In this case,
a density ``spike" with a profile $\rho \propto 1/r^{1.5}$ results, similar again to our inflow
model. 

The derived accretion rate for a steady flow is proportional to the square of the black hole mass and to
the dark matter phase space density indicator $Q=\rho_{\infty}/\sigma_{\infty}^3$.  
For black holes hosted by typical dark halos, the present accretion rate is about five orders of magnitude
higher than the usual rate derived by using the capture cross section for non-relativistic
and non-interacting particles.

Numerical applications of the derived formula were made by computing the growth of black holes
inside halos issued from cosmological simulations. We found that dark matter contributes to no more than
10\% of the total mass accreted by black hole seeds. This result is consistent with the fact that
the time integrated bolometric luminosity of quasars gives a present black hole mass density 
comparable with other independent estimates, indicating that most of the accreted mass by
seeds is baryonic in origin.

\end{document}